\documentstyle[12pt,epsfig,cite]{article}

 \newcommand{\be}{\begin{eqnarray}}
 \newcommand{\ee}{\end{eqnarray}}
 \newcommand{\beq}{\begin{equation}}
 \newcommand{\eeq}{\end{equation}}
 \newcommand{\ba}{\begin{array}{1}}
 \newcommand{\ea}{\end{array}}
 \newcommand{\bb}{\begin{thebibliography}}
 \newcommand{\eb}{\end{thebibliography}}

\title{Transition between soft physics at LHC and \\low-$x$ physics at HERA}

\author{A.A.~Grinyuk$^1$, A.V.~Lipatov$^2$, G.I.~Lykasov$^1$, N.P.~Zotov$^2$}

\begin{document}

\maketitle

\begin{center}

{\it $^1$Joint Institute for Nuclear Research, Dubna 141980, Moscow region, Russia\\

\vspace{0.1cm}
 
$^2$Skobeltsyn Institute of Nuclear Physics, Lomonosov Moscow State University, Moscow 119991, Russia}
     
\end{center}

\vspace{0.5cm}

\begin{center}

{\bf Abstract}

\end{center}

We find out the connection between the unintegrated  gluon distribution at 
low intrinsic transverse momenta and the inclusive spectrum 
of the hadrons produced in $pp$ collision
at LHC energies in the mid-rapidity region 
and low hadron transverse momenta. The parameters of this 
distribution are found from the best description of the LHC data. 
Its application to the analysis of $ep$ deep inelastic scattering 
allows us to obtain the results which describe reasonably well the H1 and ZEUS data on 
the structure functions at low $x$. A connection 
between the soft processes at LHC and small $x$ physics at HERA has been found.  

\vspace{1.0cm}

\noindent
PACS number(s): 12.38.-t, 13.85.-t

\vspace{0.5cm}

\section{Introduction}
\label{1}
\indent

Hard processes involving incoming protons, such as deep-inelastic 
lepton-proton scattering (DIS), are described using the scale-dependent parton density functions. 
Usually, these quantities are 
calculated as a function of the Bjorken variable $x$
and the square of the four-momentum transfer $q^2=-Q^2$
within the framework of the DGLAP evolution equations\cite{DGLAP} based on the 
standard collinear QCD factorization.
However, for semi-inclusive processes (such as inclusive jet production in DIS, 
electroweak boson production\cite{Ryskin:2001}, etc.) at high energies which are sensitive to the details of the parton kinematics it is more appropriate to use the parton distributions unintegrated over the transverse momentum $k_t$ or, transverse momentum depend (TMD) distributions, in the framework of the $k_t$-factorization QCD 
approach\footnote{See, for example, reviews\cite{Andersson:02} for more information.}\cite{kT}. 
The $k_t$-factorization formalism is based on the BFKL\cite{BFKL} or CCFM\cite{CCFM} evolution equations and 
provides a solid theoretical ground for the effects of initial gluon radiation. 
The unintegrated gluon $g(x,k_t)$ (u.g.d.) and 
quark $q(x,k_t)$ distributions (u.q.d.) are widely discussed and applied in phenomenological calculations in the framework
of the $k_t$-factorization QCD approach and can be found, for example, in\cite{GBW:98,GBW:99,Jung:04,Ryskin:2010,Kochelev:1998,Ball,Ermol1:11,KLZ:02,KLZ:03,KLZ:04,BLZ1,LZ1,LMZ1,HKLZ,Zotov:12,Szcz:11, Kutak:12}\footnote{The theoretical analysis of TMD was done recently in\cite{Collins,Avsar,Aybat,HHJ}.}.
In\cite{Ryskin:2001,Ryskin:2010} the unintegrated parton distributions (u.p.d.) were obtained using the so-called KMR 
prescription within the leading order (LO) 
and next-to-leading order of QCD (NLO) at large $Q^2$ 
from the known (DGLAP-evolved\cite{{DGLAP}}) parton densities determined from the global data analysis. 
These u.p.d. were successfully applied to analyze the DIS data at low $x$ and a number of processes studied at the Tevatron and LHC (see, for example,\cite{KLZ:02,KLZ:03,KLZ:04,BLZ1,LZ1,LMZ1,HKLZ,Zotov:12,Szcz:11}).
However, at small values of $Q^2$ the nonperturbative effects should be included to evaluate these distributions.
The nonperturbative effects can arise from the complex structure of the QCD vacuum. For example, 
within the instanton approach  the very fast increase of the unintegrated gluon distribution 
function at $0\le k_t\le 0.5$ GeV$/c$ and $Q^2=1$ (GeV$/c$)$^2$ is obtained\cite{Kochelev:1998}.
These results stimulated us to assume, that the u.g.d. in the proton can be determined also in 
the soft hadron production in $pp$ collisions.
 
In this paper we analyze inclusive spectra of the hadrons produced in $pp$ collisions 
at LHC energies in the mid-rapidity region 
in a context including the possible creation of soft gluons in the proton. 
We estimate the u.g.d. function at low intrinsic transverse
momenta $k_t\leq 1.5-1.6$ GeV$/$c and extract its parameters from the best description of the
LHC data at low transverse momenta $p_t$ of the produced hadrons. We also show that our u.g.d.
similar to the u.g.d. obtained in\cite{GBW:99} at large $k_t$ and different from it at low $k_t$.
The u.g.d. is directly related to the dipole-nucleon cross section within the model 
proposed in\cite{GBW:98,GBW:99} (see also\cite{Jung:04,NNN,Ivan_Nikol:02,NemNik,KR,Levin:1998}) 
which is saturated at low $Q$
or large transverse distances $r\sim 1/Q$ between quark $q$ and antiquark ${\bar q}$ in the 
$q{\bar q}$ dipole created from the splitting of the virtual photon $\gamma^*$ in the $ep$ DIS.
Here we find  a new parametrization for this dipole-nucleon cross section, as a function of
$r$, using the saturation behavior of the gluon density.

The paper is organized as following. In Section 2 we study the inclusive spectra of hadrons in $pp$ collisions and obtain
the {\it modified} u.g.d. In Section 3 we discuss the connection between the u.g.d. and the dipole cross section.
In Section 4 we apply the {\it modified} u.g.d. to describe the HERA data of the DIS structure functions:
the longitudinal ($F_L$), charm ($F_2^c$) and bottom ($F_{2}^b$) structure functions (SF).

\section{Inclusive spectra of hadrons in $pp$ collisions}
\label{sec:1}
\subsection{Unintegrated gluon distributions}
\noindent

The u.p.d. in a proton 
are a subject of intensive studies, and various approaches to 
investigate these quantities have been proposed.
At asymptotically large energies (or very small $x$) the theoretically correct 
description is given by the BFKL evolution equation\cite{BFKL} where the leading $\ln(1/x)$ contributions are 
taken into account in all orders. 
Another approach, valid for both small and large $x$, is given by the CCFM  
gluon evolution equation\cite{CCFM}. It introduces angular ordering of emissions to treat the 
gluon coherence effects correctly. 
In the limit of asymptotic high energies, it is almost equivalent to BFKL\cite{BFKL}, 
but also similar to the DGLAP evolution for large $x \sim 1$. The resulting u.g.d.
depends on two scales, the additional scale $\bar q$ is a variable 
related to the maximum angle allowed in the emission and plays the role of the evolution 
scale $\mu$ in the collinear parton densities.
In the two-scale u.p.d. obtained
from the conventional ones using the Kimber-Martin-Ryskin (KMR) prescription\cite{Ryskin:2001,Ryskin:2010},
the $k_t$ dependence in the unintegrated parton distributions enters only in the last step
of evolution. Such a procedure 
is expected to include the main part of the collinear higher-order QCD corrections.
Finally, a simple parametrization of the unintegrated gluon density was obtained 
within the color-dipole approach in\cite{GBW:98,GBW:99} on the assumption 
of saturation of the gluon density at low $Q^2$ which successfully 
described both inclusive and diffraction $ep$ scattering. This
gluon density $xg(x,k_t^2, Q_0^2)$ is given by\cite{GBW:99,Jung:04}
\begin{eqnarray}
xg(x,k_t,Q_0)=
\frac{3\sigma_0}{4\pi^2\alpha_s(Q_0)}R_0^2(x) k_t^2
\exp\left(-R_0^2(x)k^2_t\right), 
~R_0(x) = \frac{1}{Q_0}\left(\frac{x}{x_0}\right)^{\lambda/2},
\label{def:GBWgl}
\end{eqnarray}

\noindent
where $\sigma_0 = 29.12$~mb, $\alpha_s = 0.2$, $Q_0 = 1$~GeV, $\lambda = 0.277$ and
$x_0 = 4.1 \cdot 10^{-5}$.   
This simple expression corresponds 
to the Gaussian form for the effective dipole cross section ${\hat\sigma}(x,r)$
as a function of $x$ and the relative transverse separation ${\bf r}$ of the $q{\bar q}$ pair\cite{GBW:99}. 
In fact, this form can be more complicated. In this paper we study this point and try to find a 
parametrization   
for $xg(x,k_t,Q_0)$, which is related to ${\hat\sigma}(x,r)$, from the best description of the 
inclusive spectra 
of charge hadrons produced in $pp$ collisions at LHC energies and mid-rapidity region.      

\subsection{Quark-gluon string model (QGSM) including gluons}
\noindent

The soft hadron production in $pp$ collisions at not too large momentum transfer 
can be analyzed within the soft QCD models, namely, the quark-gluon 
string model (QGSM)\cite{kaid1,LS:1992,BLL:2010}
or the dual parton model (DPM)\cite{capell2}. The cut $n$-pomeron 
graphs calculated within these models result in a reasonable description at small but nonzero rapidities. 
However, it has been shown recently\cite{BGLP:2011,BGLP:2012} that there are some 
difficulties in using the QGSM to analyze inclusive spectra in $pp$ collisions 
in the mid-rapidity region and at the initial energies above the ISR one. 
However, it is due to the Abramovsky-Gribov-Kancheli cutting rules (AGK)\cite{AGK} at mid-rapidity 
($y\simeq 0$), when only one-pomeron Mueller-Kancheli diagrams contribute to the inclusive spectrum 
$\rho_h(y\simeq 0, p_t)$. 
To overcome these difficulties it was assumed\cite{BGLP:2011} that there are soft gluons or the so called
{\it intrinsic} gluons in the proton suggested in\cite{Brodsky:1981}, which split  
into $q{\bar q}$ pairs and should vanish at the zero intrinsic transverse momentum ($k_t\sim 0$)
because at $k_t\sim 0$ the conventional QGSM (without the {\it intrinsic} gluons)\cite{kaid1} 
is applied very well. 
The inclusive spectrum of hadrons $Ed\sigma/d^3p\,(y\simeq 0)\equiv\rho_h(y\simeq 0, p_t)$ was split into
two parts, the quark contribution $\rho_q(y\simeq 0, p_t)$ and the gluon one 
and their energy dependence was calculated in\cite{BGLP:2011,BGLP:2012}
\begin{eqnarray}
\rho_h(y\simeq 0,p_t)=\rho_q(y\simeq 0,p_t)+\rho_g(y\simeq 0,p_t),
\label{def:rhotot}
\end{eqnarray}

\noindent
where $\rho_q(y\simeq 0,p_t)$ is the quark contribution and $\rho_g(y\simeq 0,p_t)$ is the contribution of gluons to the 
spectrum $\rho_h(y\simeq 0,p_t)$. It was shown\cite{BGLP:2012} that the AGK cutting rules for the inclusive spectrum
at $x\simeq 0$ can be proofed within the QGSM\cite{kaid1} and $\rho_q(y\simeq 0,p_t)$ can be presented in the form
\be
\rho_q(0,p_t)=
\sum_{n=1}^\infty \sigma_n(s)\phi_n^q(0,p_t),
\label{def:invspqn}
\ee

\noindent
where $\phi_n^q$ is the convolution of the quark distribution and the fragmentation function of the quark $q$ to the
hadron $h$ that at $x\simeq 0$ $(y\simeq 0)$ is proportional to the pomeron number $n$, see~(9) and (10) 
in\cite{BGLP:2012}. Therefore, (\ref{def:invspqn}) is presented in the form 
\begin{eqnarray}
\rho_q(0,p_t)=
{\tilde\phi}_q(0,p_t)\sum_{n=1}^\infty n \sigma_n(s),
\label{def:invspqnn}
\end{eqnarray}

\noindent
where $\sigma_n$ is the cross section for production of the $n$-pomeron chain (or $2n$ quark-antiquark strings) 
decaying into hadrons, calculated within the ``eikonal approximation''\cite{ter-mar},
the function ${\tilde\phi}_q(0,p_t)$ is related to the pomeron-hadron vertex in the Mueller-Kancheli diagram. 
Inserting $\sigma_n$ into ({\ref{def:invspqnn}) the quark contribution $\rho_q(0,p_t)$ of the spectrum is presented in 
the form\cite{BGLP:2012}
\be
\rho_q(0,p_t)=
g(s/s_0)^{\Delta}{\tilde\phi}_q(0,p_t),
\label{def:invspq}
\ee

\noindent
where $g=21$ mb, $\Delta=\alpha_P(0)-1\simeq 0.12$, $\alpha_P(0)$ is the intercept of the sub-critical Pomeron, 
the function
${\tilde\phi}_q(0,p_t)$ is found from the fit of the SPS and LHC data on the inclusive spectra of charged hadrons in
the mid-rapidity at the initial energies from 540 GeV till 7 TeV.

The {\it intrinsic} gluons split into the sea $q{\bar q}$ pairs, therefore its contribution to the inclusive spectrum
is due to the contribution of the sea quarks, which, according to the QGSM ideology\cite{kaid1}, contribute to the $n$-pomeron shower at $n\geq$ 2. 
Therefore, the contribution $\rho_g(y\simeq 0,p_t)$ is presented in the form similar to (\ref{def:invspqnn}) replacing ${\tilde\phi}_q(0,p_t)$ by $\tilde{\phi}_g(0,p_t)$ and inputting $n\geq 2$
\be
\displaystyle \rho_g(0,p_t)=\tilde{\phi}_g(0,p_t)\sum_{n=2}^\infty n\sigma_n(s)\equiv \atop {
\displaystyle \tilde{\phi}_g(0,p_t)\left(\sum_{n=1}^\infty n\sigma_n(s)-\sum_{n=1}^\infty \sigma_n(s)\right) = \atop {
\displaystyle {\tilde\phi}_g(0,p_t)(g(s/s_0)^{\Delta}-\sigma_{nd}),}}
\label{def:invspg}
\ee

\noindent
where $\sigma_{nd}=\sum_{n=1}^\infty \sigma_n(s)$ is the nondiffractive cross section, the function ${\tilde\phi}_g(0,p_t$
is also found from the fit of the SPS and LHC data on the inclusive spectra of charged hadrons in the mid-rapidity
region\cite{BGLP:2012}. The following parametrizations for ${\tilde\phi}_q(0,p_t)$ and
 ${\tilde\phi}_g(0,p_t)$ were found\cite{BGLP:2011}:   
\begin{eqnarray}
\displaystyle {\tilde\phi}_q(0,p_t)=A_q\exp(-b_q p_t),\atop {
\displaystyle {\tilde\phi}_g(0,p_t)=A_g\sqrt{p_t}\exp(-b_g p_t),}
\label{def:phiq}
\end{eqnarray}

\begin{figure}[h!!]
\begin{center}
\epsfig{file=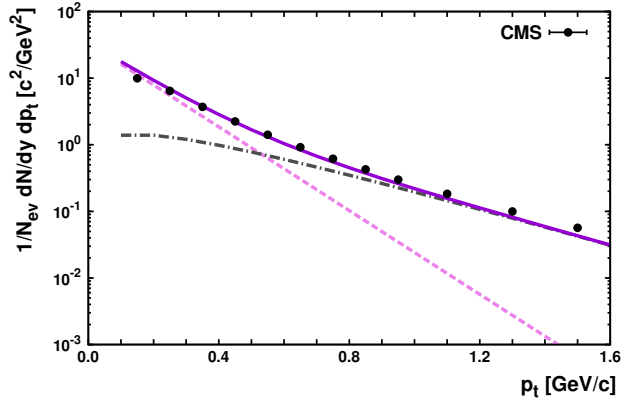,width=8.2cm}
\end{center}
\caption[Fig.1]{The inclusive spectrum of the charged hadron as a function of $p_t$ (GeV/c)
in the central rapidity region ($y=0$) at $\sqrt{s}=7$ TeV 
at $p_t\leq 1.6$ GeV/c compared with the CMS measurements\cite{CMS} which are very close to the
ATLAS data\cite{ATLAS}. The dashed and dash-dotted curves correspond to the quark and 
gluon contribution given by (\ref{def:invspq}) and (\ref{def:invspg}), respectively. The 
solid curve represents their sum.}
\label{Fig_Soft}
\end{figure} 

\begin{figure}[h!!]
\epsfig{file=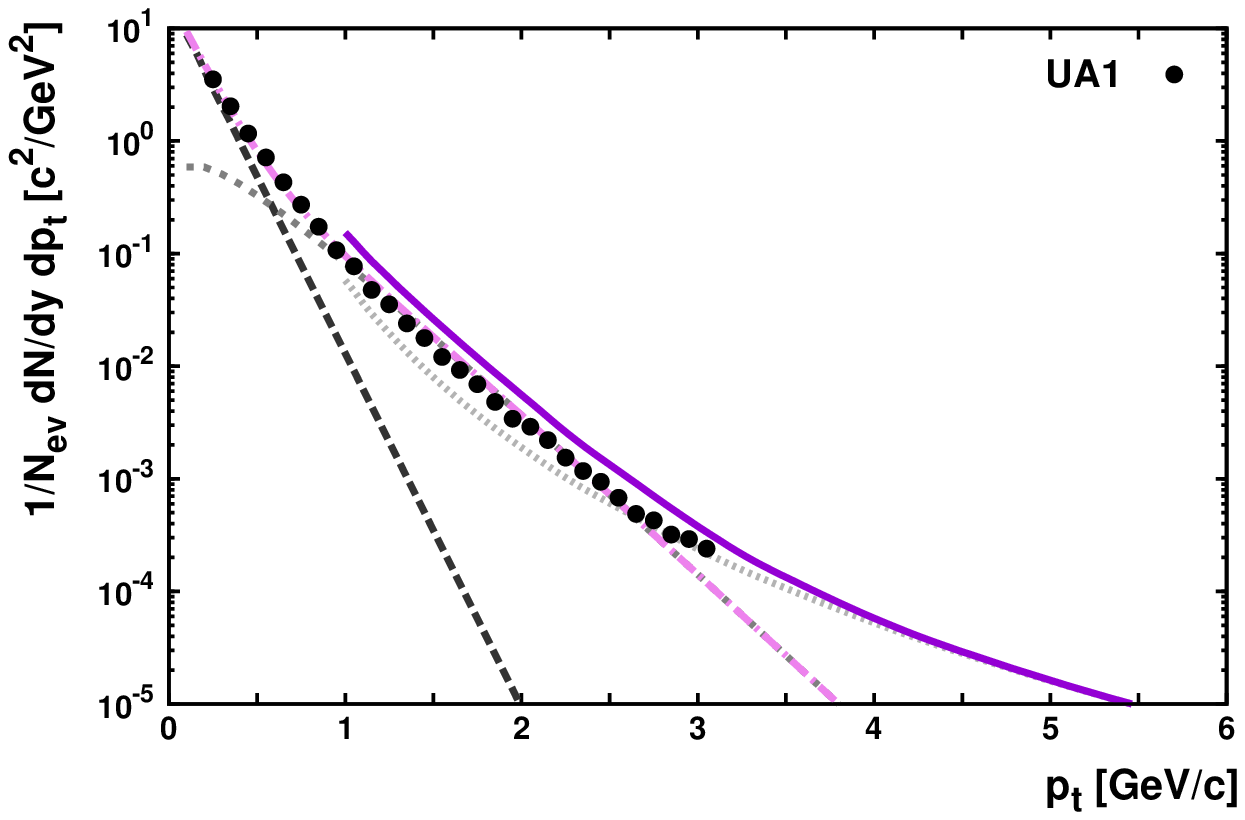,width=8.2cm}
\epsfig{file=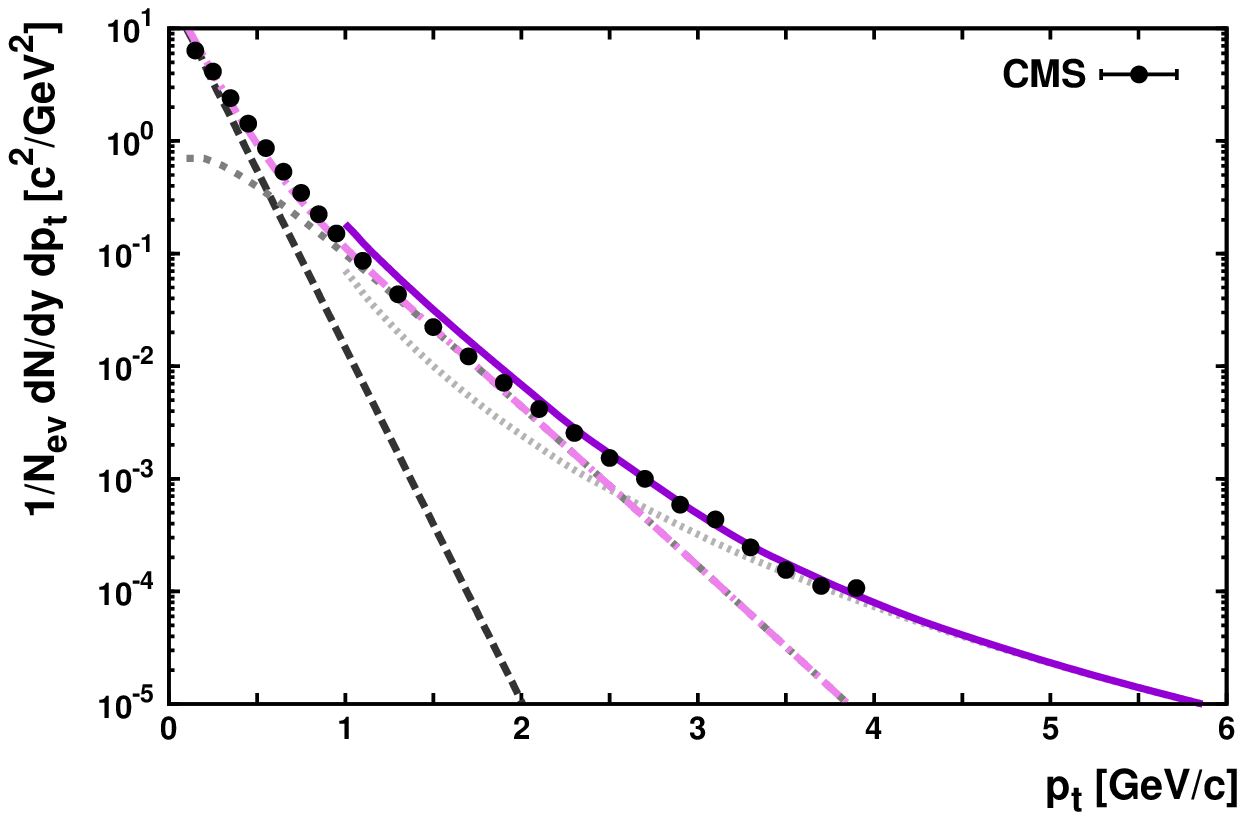,width=8.2cm}
\caption{The inclusive spectrum of charged hadron as a function of $p_t$ (GeV/c)
in the central rapidity region ($y=0$) at $\sqrt{s}= 540$ GeV (left) and 
$\sqrt{s} = 900$ GeV (right) compared with the UA1\cite{UA1} and CMS\cite{CMS} data. 
The long dashed curves are the quark contribution $\rho_q(x=0,p_t)$
(\ref{def:invspq}), the short dashed curves correspond to the gluon one $\rho_g(x=0,p_t)$ (\ref{def:invspg}), the dash-dotted curves are the sum of the quark and gluon contributions (\ref{def:rhotot}), the dotted curves correspond to the perturbative LO QCD\cite{BGLP:2012}. 
The solid curves represent the sum of the calculations within the soft QCD including the gluon 
contribution (\ref{def:rhotot}) and the perturbative LO QCD.} 
\label{Fig_2}
\end{figure}

\begin{figure}[h!]
\epsfig{file=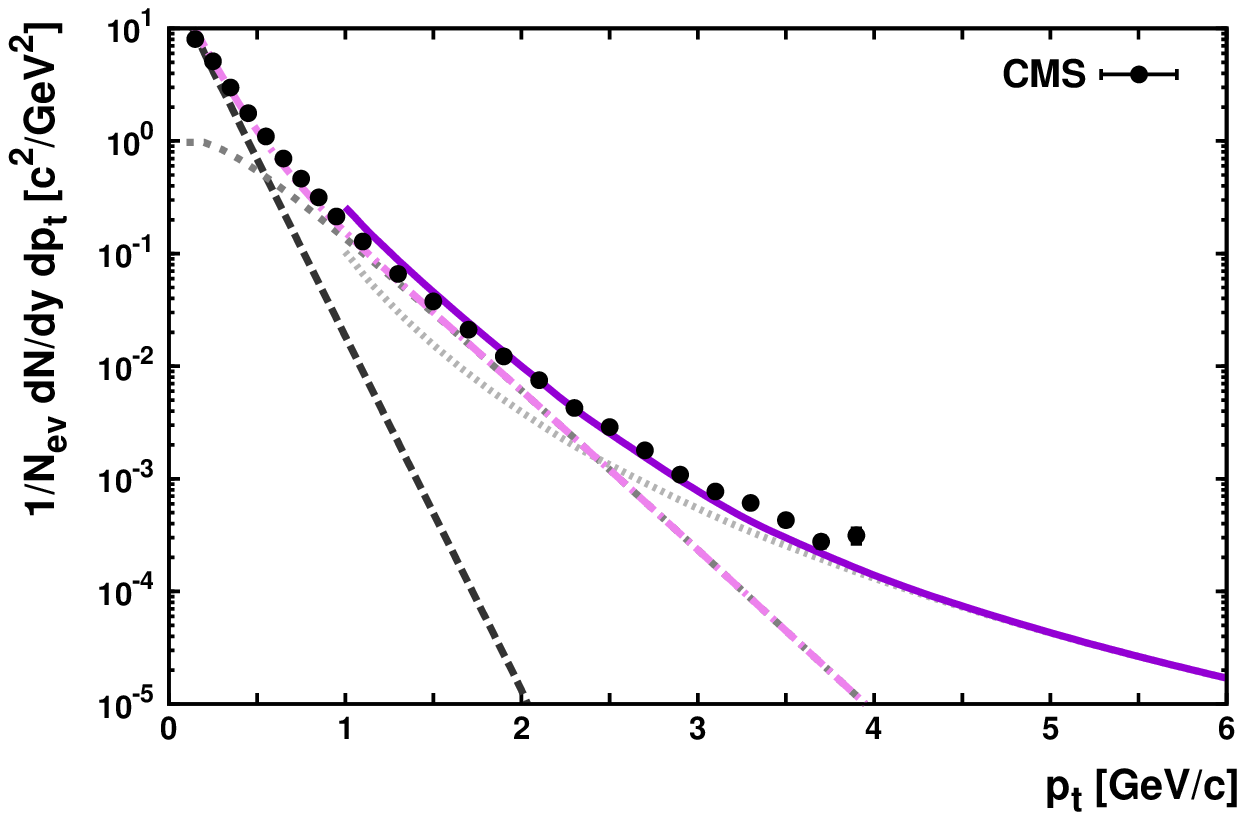,width=8.2cm}
\epsfig{file=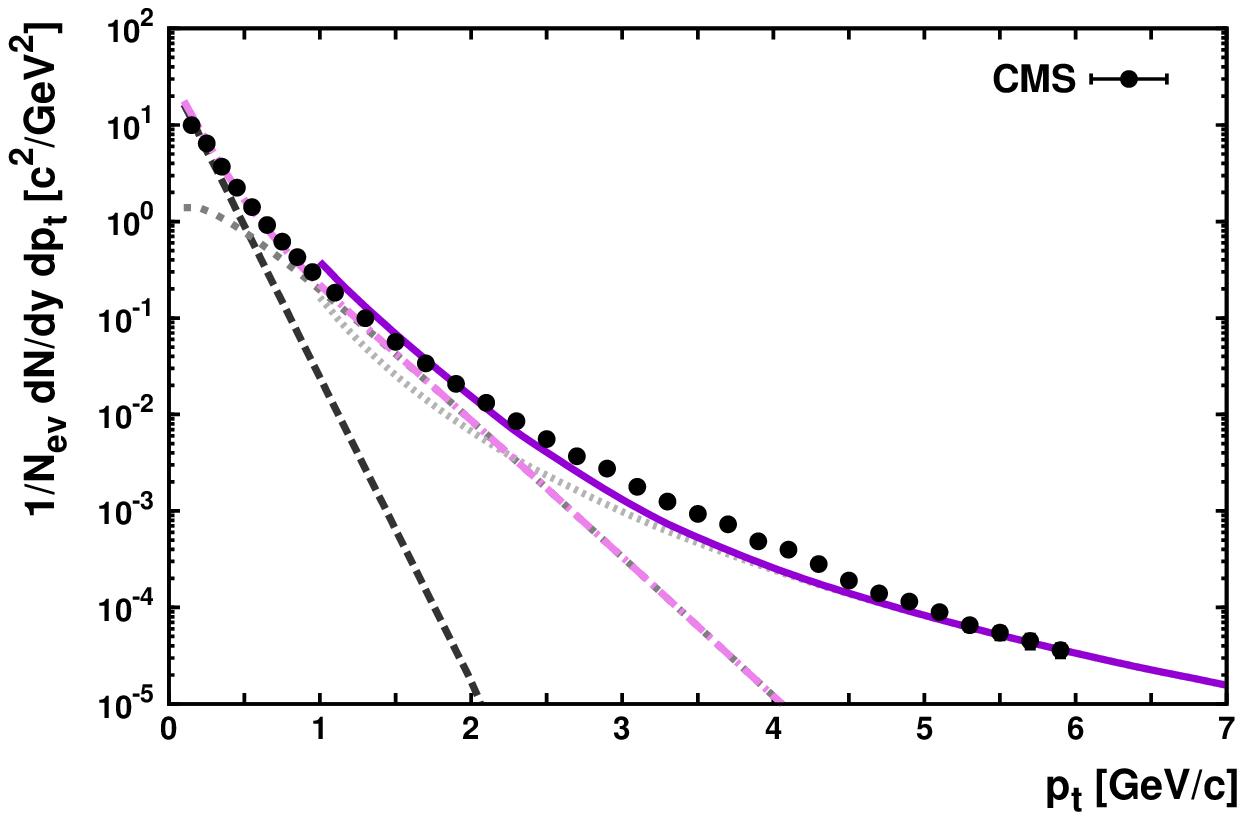,width=8.2cm}
\caption{The inclusive spectrum of charged hadron as a function of $p_t$ (GeV/c)
in the central rapidity region ($y=0$) at $\sqrt{s}=2.36$ TeV (left) and $\sqrt{s}=7$ TeV (right)
compared with the CMS\cite{CMS} and ATLAS\cite{ATLAS} data. 
Notation of all curves is the same as in Fig.~2.} 
\label{Fig_3}
\end{figure}

\noindent
where $s_0=1$ GeV$^2$, $g=21$ mb, $\Delta=0.12$.
The parameters are fixed\cite{BGLP:2011} from the fit to the data on the $p_t$ distribution of
charged particles at $y=0$:
$A_q=4.78\pm 0.16$ (GeV$/$c)$^{-2}$,~$b_q=7.24\pm 0.11$ (GeV/c)$^{-1}$ and
 $A_g=1.42\pm 0.05$ (GeV$/$c)$^{-2}$;~ 
$b_g=3.46\pm 0.02$ (GeV/c)$^{-1}$. 
In Fig.~\ref{Fig_Soft} we
illustrate the fit of the inclusive spectrum of charged hadrons produced in 
$pp$ collisions at $\sqrt{s}=7$ TeV and the central rapidity region at the hadron transverse momenta
$p_t\leq 1.6$ GeV/c. Here the solid line corresponds to the quark contribution $\rho_q$,  
the dashed line is the gluon contribution $\rho_g$, and the dotted curve is the sum of these contributions
$\rho_h$ given by (\ref{def:rhotot}). The little discrepancy between the data and our calculation (the dotted line) 
at $p_t> 1.2$ GeV/c disappears if the contribution of the perturbaive QCD within the LO is included. It is shown
in Figs.~\ref{Fig_2} and \ref{Fig_3}, where the inclusive hadron spectrum is presented 
at $\sqrt{s}= 0.54$, $0.9$, $2.36$ and $7$ TeV, where the dash-dotted line is the sum of the spectrum $\rho_h(x=0,p_t)$ (\ref{def:rhotot}) and the short dash-dotted line is the result of calculation within the LO PQCD\cite{BGLP:2012}. 
The quite satisfactory description of the data on such spectra was obtained 
in\cite{BGLP:2012} at the SPS and LHC energies using (\ref{def:rhotot}) for $\rho_h(x=0,p_t)$ and the LO PQCD calculations both together.
Therefore we conclude that the energy dependence of the inclusve spectrum of charged hadrons
produced in the $pp$ collision at the mid-rapidity region is reasonably well described using the parametrization (\ref{def:phiq}) for ${\tilde\phi}_q(0,p_t)$ and ${\tilde\phi}_g(0,p_t)$.
  
\subsection{Modified unintegrated gluon distributions}
\noindent

As it can be seen in Figs.~\ref{Fig_Soft} --- \ref{Fig_3} the contribution to the inclusive spectrum at $y\simeq 0$ due to
the {\it intrinsic} gluons is sizable at low $p_t< 2$ GeV/c, e.g., in the soft kinematical region. Therefore,
we can estimate this contribution within the nonperturbative QCD model, similar to the QGSM\cite{kaid1}.   
We calculate the gluon contribution ${\tilde\phi}_g(x\simeq 0,p_t)$ entering into 
(\ref{def:invspg}) as the cut graph (Fig.~\ref{Fig_Pomeron}, right) of the one-pomeron exchange in 
the gluon-gluon interaction (Fig.~\ref{Fig_Pomeron}, left) 
using the splitting of the gluons into the $q{\bar q}$ pair.  
The right diagram of Fig.~\ref{Fig_Pomeron} corresponds to the creation of two colorless
strings between the quark/antiquark $(q/{\bar q})$ and antiquark/quark $({\bar q}/q)$. Then,
after their brake, $q{\bar q}$ are produced and fragmented into the hadron $h$.   
Actually, the calculation can be  
made in a way similar to the calculation of the sea quark contribution to the inclusive 
spectrum within the QGSM\cite{kaid1}, e.g., the contribution ${\tilde\phi}_g(0,p_t)$ is 
presented as the sum of two convolution functions
\begin{eqnarray}
{\tilde\phi}_g(x,p_t)=F_q(x_+,p_{ht})F_{\bar q}(x_-,p_{ht})+F_{\bar q}(x_+,p_{ht})F_q(x_-,p_{ht}),
\label{def:rhog}
\end{eqnarray}

\noindent
where the function $F_{q({\bar q})}(x_+,p_{ht})$ corresponds to the production of the hadron $h$
from the decay of the upper vertex of $q{\bar q}$ string and $F_{q({\bar q})}(x_-,p_{ht})$ corresponds 
to the production of $h$ from the decay of the bottom vertex of $q{\bar q}$ string. 
They are calculated as the following convolution:
\begin{eqnarray}
F_{q({\bar q})}(x_\pm,p_{ht})=
\int_{x\pm}^1dx_1
\int d^2k_{1t}\,f_{q({\bar q})}(x_1,k_{1t})\,G_{q({\bar q})\rightarrow h}
\left(\frac{x_\pm}{x_1},\,p_{ht}-k_{1t})\right).
\label{def:Fqbrq}
\end{eqnarray}

\begin{figure}[h!]
\centerline{\includegraphics[width=0.8\textwidth]{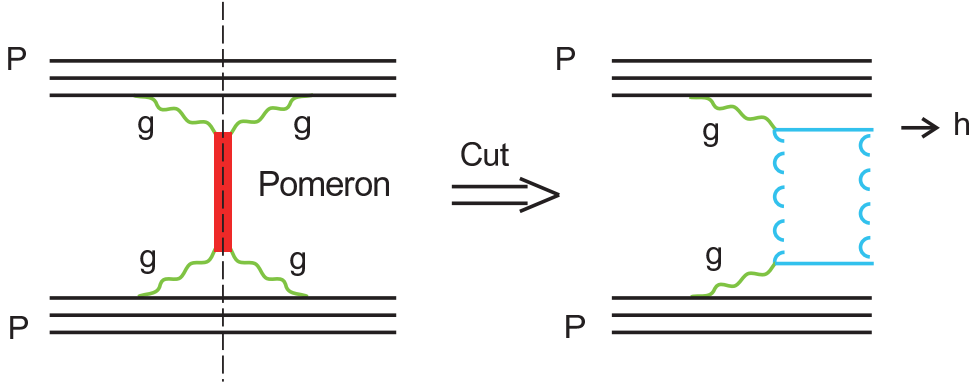}}
\caption{The one-pomeron exchange graph between two gluons in the elastic $pp$ scattering (left) 
and the cut one-pomeron due to the creation of two colorless strings between 
quarks/antiquarks that decay into $q{\bar q}$ pairs, 
which are drawn as the semi-circles (right)\cite{kaid1}.} 
\label{Fig_Pomeron}
\end{figure}

\noindent
Here $G_{q({\bar q})\rightarrow h}(z,{\tilde k}_t)=zD_{q({\bar q})\rightarrow h}(z,{\tilde k}_t)$,
$D_{q({\bar q})\rightarrow h}(z,{\tilde k}_t)$ is the fragmentation function (FF) of the quark (antiquark)
to the hadron $h$, $z=x_\pm/x_1,{\tilde k}_t=p_{ht}-k_t$, 
$x_{\pm}=0.5(\sqrt{x^2+x_t^2}\pm x)$, $x_t=2\sqrt{(m_h^2+p_t^2)/s}$.
At $x\simeq 0$, we get that $x_+$ and $x_-$ are equivalent to each other, e.g., $x_+=x_-=m_t/\sqrt{s}$. 
The distribution of sea quarks (antiquark) 
$f_{q({\bar q})}$ is related to the splitting function ${\cal P}_{g\rightarrow q{\bar q}}$ of gluons to 
$q{\bar q}$ by
\begin{eqnarray}
f_{q({\bar q})}(z,k_t)=\int_z^1 g(z_1,k_t,Q_0){\cal P}_{g\rightarrow q{\bar q}}
\left(\frac{z}{z_1}\right)\frac{dz_1}{z_1}~,
\label{def:fqbq}
\end{eqnarray}

\noindent
where $g(z_1,k_{1t},Q_0)$ is the u.g.d. The gluon splitting function ${\cal P}_{g\rightarrow q{\bar q}}$
was calculated within the Born approximation. 
In (\ref{def:fqbq}) we assumed the collinear splitting of the {\it intrinsic} gluon to the $q{\bar q}$ pair
because values of $k_t$ are not zero but small.

Calculating the diagram of Fig.~4 (right) by the use of (\ref{def:rhog}) --- (\ref{def:fqbq}) for the gluon 
contribution $\rho_g$ we took the FF to charged hadrons, pions, kaons, and $p{\bar p}$ pairs obtained
within the QGSM\cite{Shabelsk:1992}. From the best description of $\rho_g(x\simeq 0,p_{ht})$, see
its parametrization given by~(\ref{def:phiq}), we found the form     
for the $xg(x,k_t,Q_0)$ which was fitted in the following way:
\begin{eqnarray}
\displaystyle xg(x,k_t,Q_0)=\frac{3\sigma_0}{4\pi^2\alpha_s(Q_0)} C_1 (1-x)^{b_g}\times \atop {
\displaystyle \times \left(R_0^2(x)k_t^2+C_2(R_0(x)k_t)^a\right) \exp\left[-R_0(x)k_t-d(R_0(x)k_t)^3\right],}
\label{def:gldistrnew}
\end{eqnarray}

\noindent
where $R_0(x)$ is defined in (1). The coefficient $C_1$ was found from the following normalization:
\begin{eqnarray}
g(x,Q_0^2)=\int_0^{Q_0^2} dk_t^2 \, g(x,k_t^2,Q_0^2),
\label{def:BFKL}
\end{eqnarray}

\noindent
and the parameters 
$a=0.7$, $C_2\simeq 2.3$, $\lambda=0.22$, $b_g=12$, $d=0.2$, $C_3=0.3295$
were found from the best fit of the LHC data on the inclusive spectrum of charged
hadrons produced in $pp$ collisions and in the mid-rapidity region as it can be seen in 
Figs.~\ref{Fig_Soft} --- \ref{Fig_3}. 

\begin{figure}[h!]
\begin{center}
\epsfig{file=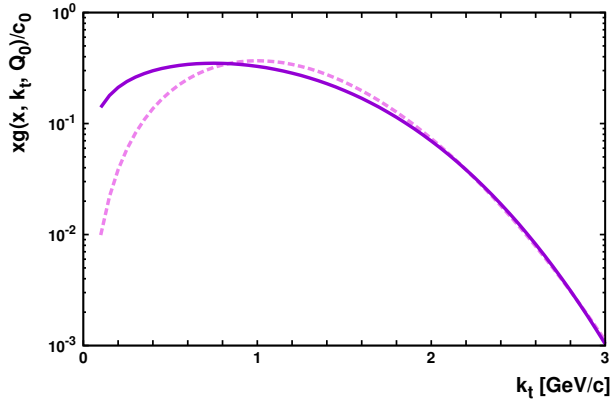,width=8.2cm}
\end{center}
\caption 
{The unintegrated gluon distribution $xg(x,k_t,Q_0)/c_0$ 
(where $c_0=3\sigma_0/(4\pi^2\alpha_s(Q_0))$) as a function of $k_t$
at $x=x_0$ and $Q_0=1$ GeV/c. The solid and dashed curves correspond 
to the the modified u.g.d. (\ref{def:gldistrnew}) and original GBW 
gluon density\cite{GBW:98} given by (\ref{def:GBWgl}), respectively.} 
\label{Fig_4}
\end{figure} 

In Fig.~5 we present the modified u.g.d. obtained by calculating the cut one-pomeron graph of Fig.~4
and the original GBW u.g.d.\cite{GBW:98} as a function of the transverse gluon momentum
$k_t$. One can see that the modified u.g.d.
(the solid line in Fig.~5) is different from the original GBW gluon density\cite{GBW:98} at small
$k_t~<~1.5$ GeV/c and coincides with it at larger $k_t$. This is due to the sizeble 
contribution of $\rho_g$ in~(\ref{def:invspg}) and (\ref{def:phiq}) 
to the inclusive spectrum $\rho(p_t)$ of charged hadrons produced
in $pp$ collisions at LHC energies and in the mid-rapidity region (see the dashed line in Fig.~1). 

Let us also note that, as it was shown recently in\cite{GJLLZ:2012}, the modified GBW given by 
(\ref{def:gldistrnew}) describes reasonably well the HERA data on the proton longitudinal structure function 
$F_L(Q^2)$ at low $x$.
\section{Saturation dynamics} \indent

According to\cite{GBW:98,GBW:99} (see also\cite{Ivan_Nikol:02,NemNik,KR}), the u.g.d. can be related to the cross section
${\hat\sigma}(x,r)$ of the 
$q{\bar q}$ dipole with the nucleon. This dipole is created from the split of the virtual exchanged photon
$\gamma^*$ to $q{\bar q}$ pair in $ep$ deep inelastic scattering (DIS).The relation at the fixed 
$Q_0^2$ is the following\cite{GBW:99}:
\be
{\hat\sigma}(x,r) = \frac{4\pi\alpha_s(Q_0^2)}{3}\int\frac{d^2k_t}{k_t^2}
\left\{ 1-J_0(rk_t)\right\}xg(x,k_t).
\label{def:sigxr}
\ee  

\noindent 
Using the simple form for $xg(x,k_t)$ given by (\ref{def:GBWgl}) as input to (\ref{def:sigxr}) 
on can get the following form for the dipole cross section:
\be
{\hat\sigma}_{\rm GBW}(x,r)=\sigma_0\left\{1-\exp\left(-\frac{r^2}{4R_0^2(x)}\right)\right\}.
\label{def:sigGBW}
\ee

\noindent
However, the modified u.g.d. given by (\ref{def:gldistrnew}) inputted to (\ref{def:sigxr})
results in a more complicated form for ${\hat\sigma}(x,r)$:
\be
{\hat\sigma}_{\rm modif}(x,r)=\sigma_0\left\{1-\exp\left(-\frac{b_1r}{R_0(x)}-\frac{b_2r^2}{R_0^2(x)}\right)\right\},
\label{def:sigxrmod}
\ee

\noindent
where $b_1=0.045$ and $b_2=0.3$.
In Fig.~6 we show the difference between the dipole cross section ${\hat\sigma}_{\rm GBW}(x=x_0,r)$\cite{GBW:99} and  
${\hat\sigma}_{\rm modif}(x=x_0,r)$ obtained from the modified u.g.d. given by (\ref{def:gldistrnew}). 
The saturation effect 
means that the dipole cross section becomes constant when $r>2R_0$. As it was shown in\cite{GBW:98,GBW:99}, at low
$Q^2$ the transverse $\gamma^* p$ cross section calculated within the dipole model is about constant when $QR_0<1$. 
It means that $Q_s\sim 1/R_0$ can be treated as the saturation scale. One can see in Fig.~6 that the 
modified dipole 
cross section is saturated a little earlier than the GBW one.  

\begin{figure}[h!]
\begin{center}
\epsfig{file=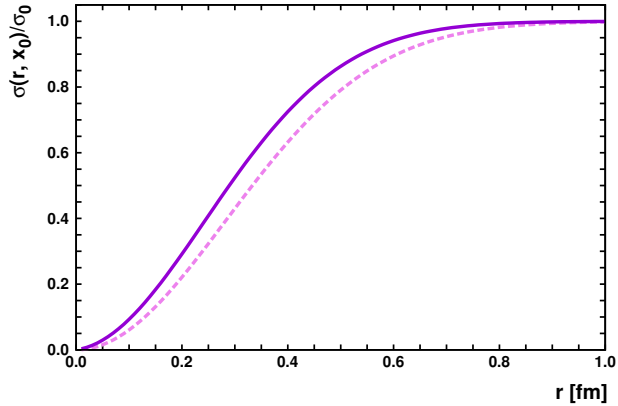,width=8.2cm}
\end{center}
\caption 
{The dipole cross section ${\hat\sigma}/\sigma_0$ at $x=x_0$ as a function of $r$.
The solid and dashed curves correspond to our calculations and 
calculation of\cite{GBW:99}, respectively.} 
\label{Fig_5}
\end{figure}

\begin{figure}[h!]
\begin{center}
\epsfig{file=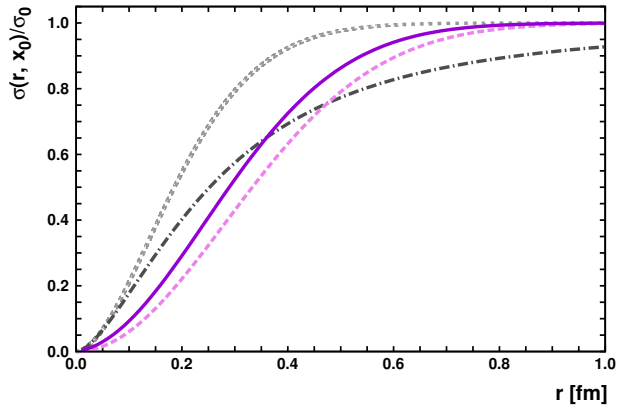,width=8.2cm}
\end{center}
\caption 
{The dipole cross section ${\hat\sigma}/\sigma_0$ at $x=x_0$ as a function of $r$.
The solid, dashed, dash-dotted and dotted curves correspond to our calculations,
calculations of\cite{GBW:99}, calculations of\cite{NNN} and\cite{Marquet:2010},
respectively.} 
\label{Fig_6}
\end{figure}

There are different forms of the dipole cross sections suggested in\cite{NNN,Ivan_Nikol:02,NemNik,Marquet:2010}. 
The dipole cross section can be presented in the general form\cite{GBW:98}:
\be
{\hat\sigma}(x,r)=\sigma_0g({\hat r}^2),
\label{def:sigdipgen}
\ee

\noindent
where ${\hat r}=r/(2R_0(x))$. The function $g({\hat r}^2)$ can be written in the form\cite{NNN} 
\be
g({\hat r}^2)={\hat r}^2\log\left(1+\frac{1}{{\hat r}^2}\right),
\label{def:sigdipNZ90}
\ee

\noindent
or in the form\cite{Marquet:2010}
\be
g({\hat r}^2)=1-\exp\left\{-{\hat r}^2\log\left(\frac{1}{\Lambda r}+e\right)\right\},
\label{def:sigdipMV98}
\ee 

\noindent
where saturation occurs for larger $r$.
In Fig.~7 we illustrate the dipole cross sections ${\hat\sigma}/\sigma_0$ at $x=x_0$ which are saturated 
at $r>0.6$ fm, obtained in\cite{NNN,McLerran:1998,Marquet:2010}.
They are compared with the results of our calculations (solid line) given by (\ref{def:sigxrmod}).  
The solid curve in Fig.~7 corresponds to the modified u.g.d. given by (\ref{def:gldistrnew}),
which allowed us to describe the LHC data on inclusive spectra of hadrons produced in
the mid-rapidity region of $pp$ collision at low $p_t$. Therefore, the form of the dipole-nucleon cross 
sections presented in Fig.~7 can be verified by the description of the last LHC data on hadron spectra in soft 
kinematical region.

Comparing the solid curve ("Modified $\sigma$") and dashed curve ("GBW $\sigma$") in Fig.~7 one can
see that ${\hat\sigma}_{\rm modif}(x,r)$ given by (\ref{def:sigxrmod}) is saturated earler than  
${\hat\sigma}_{\rm GBW}(x,r)$ given by (\ref{def:sigGBW}) with
increasing the transverse dimension $r$ of the $q{\bar q}$ dipole.
If $R_0=(1/$GeV$)\,(x/x_0)^{\lambda/2}$, according to\cite{GBW:98,GBW:99}, then the saturation scale has the form
$Q_s\sim 1/R_0=Q_{s0}(x_0/x)^{\lambda/2}$, where $Q_{s0}=1$~GeV$= 0.2$~fm$^{-1}$. The saturation of the dipole 
cross section (\ref{def:sigGBW}) sets in when $r\sim 2R_0$ or $Q_{s}\sim (Q_{s0}/2)(x_0/x)^{\lambda/2})$.
Comparing the saturation properties of the modified $\sigma$ and GBW $\sigma$ presented in Fig.~7 one 
can get slightly larger value for $Q_{s0}$ in comparison with $Q_{s0}=1$~GeV.

\section{Proton structure functions} 
\indent

The basic information on the internal structure of the proton can be extracted from the process of deep inelastic $ep$ scattering. Its differential cross-section has the form:
\be
  {d^2\sigma\over dx dy} = {2\pi \alpha_{em}^2\over x Q^4} \left[ \left(1 - y + {y^2\over 2} 
\right)F_2(x,Q^2) - {y^2\over 2}F_L(x,Q^2)\right],
\ee

\noindent
where $F_2(x,Q^2)$ and $F_L(x,Q^2)$ are the transverse and longitudinal proton structure
functions, $x = Q^2/2(p \cdot q)$ and
$y = Q^2/x s$ are the usual Bjorken variables with $p$, $q$ and $s$ being the
proton and photon four-momenta and total $ep$ center-of-mass energy, respectively.
In the present paper we will concentrate on the charm and beauty part of $F_2(x,Q^2)$
and on the longitudinal SF $F_L(x,Q^2)$.  
Theoretical analysis\cite{GMVFNS1,GMVFNS2,GMVFNS3,GMVFNS4,GMVFNS5,GMVFNS6} 
have generally  confirmed that $F_2^c(x,Q^2)$ and $F_2^b(x,Q^2)$ data 
can be described through perturbative generation of charm and 
beauty within QCD. The longitudinal SF $F_L(x,Q^2)$ is very sensitive 
to QCD processes since it is directly connected to the gluon 
content of the proton. 
It is equal to zero in the parton model with spin $1/2$ partons and has nonzero values 
in the framework of pQCD.

In the $k_t$-factorization approach\cite{kT},  
the study of charm and beauty contributions to the proton SF $F_2(x,Q^2)$
and longitudinal SF $F_L(x,Q^2)$ has been performed previously in\cite{KLZ:02,KLZ:03,KLZ:04}.
In these calculations the different approaches to evaluate the
unintegrated gluon density in a proton have been tested and a
reasonable well agreement with the HERA data has been found, in particular, 
with the CCFM-evolved gluon density proposed in\cite{Jung:04}.
Below we apply the unintegrated gluon distribution given by~(7) to describe recent experimental 
data\cite{H1,H1EL,H1ZEUS:2012,H1F2cb:2011,H1F2c:2010,ZEUSF2b:2011,ZEUSF2cb:2010}
on $F_2^c(x,Q^2)$, $F_2^b(x,Q^2)$ and $F_L(x,Q^2)$ taken by
the H1 and ZEUS collaborations at HERA. 
The main formulas have been obtained previously in\cite{KLZ:02}. Here we only recall some of them.

According to the $k_t$-factorization prescription,
the considered proton SF can be
calculated as a following double convolution:
\be
  F_2^{c,b}(x, Q^2) = \int {dy \over y} \int dk_t^2 \, {\cal C}_2(x/y,k_t^2,Q^2,\mu^2) f_g(y,k_t^2,\mu^2),
\ee
\be
  F_L(x, Q^2) = \sum_f e_f^2 \int {dy \over y} \int dk_t^2 \, {\cal C}_L(x/y,k_t^2,Q^2,\mu^2) f_g(y,k_t^2,\mu^2),
\ee

\noindent
where $e_f^2$ is the electric charge of the quark of flavor $f$. The hard coefficient 
functions ${\cal C}_{2,L}(x,k_t^2,Q^2,\mu^2)$ correspond to the quark-box diagram 
for the photon-gluon fusion subprocess and have been calculated in\cite{KLZ:02}.
Numerically, here we set charm and beauty quark masses to $m_c = 1.4$~GeV and $m_b = 4.75$~GeV
and use the LO formula for the strong coupling constant $\alpha_s(\mu^2)$ with $n_f = 4$ quark flavours
at $\Lambda_{\rm QCD} = 200$~MeV, such that $\alpha_s(M_Z^2) = 0.1232$.
Note that in order to take into account the NLO corrections
(which are important at low $Q^2$) in our numerical calculations we apply the method 
proposed in\cite{Brodsk_K:99}. Following\cite{KLZ:04,Brodsk_K:99}, we 
use the shifted value of the renormalization scale $\mu_R^2 = K \, Q^2$, where
$K = 127$. As is was shown in\cite{Brodsk_K:99}, this shifted scale in the DGLAP approach 
at LO approximation leads to the results which are very close to the NLO ones. In the case of  $k_t$-factorization this
procedure gives us a possibility to take into account additional higher-twist and non-logorithmic  NLO corrections\cite{KLZ:04}.

\begin{figure}[h!]
\begin{center}
\epsfig{file=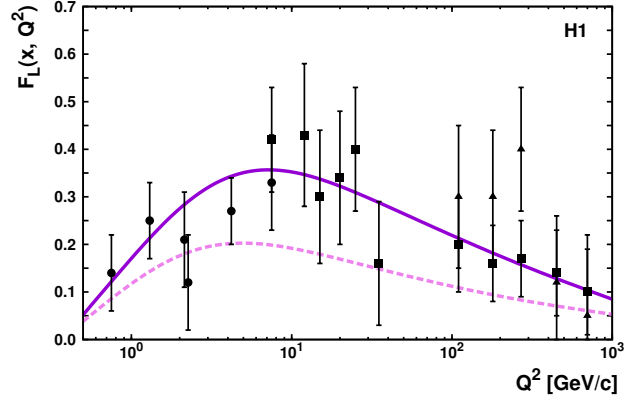,width=8.2cm}
\end{center}
\caption 
{The longitudinal structure function $F_{L}(x, Q^2)$ at fixed $W = 276$ GeV 
and $\mu_R^2 = K\cdot Q^2$,
where $K=127$\cite{Brodsk_K:99}. Notation of all curves is the same as in Fig.~5. 
The H1 data are taken from\cite{H1,H1EL}.} 
\label{Fig_7}
\end{figure} 

\begin{figure}[h!!]
\centerline{\includegraphics[width=0.8\textwidth]{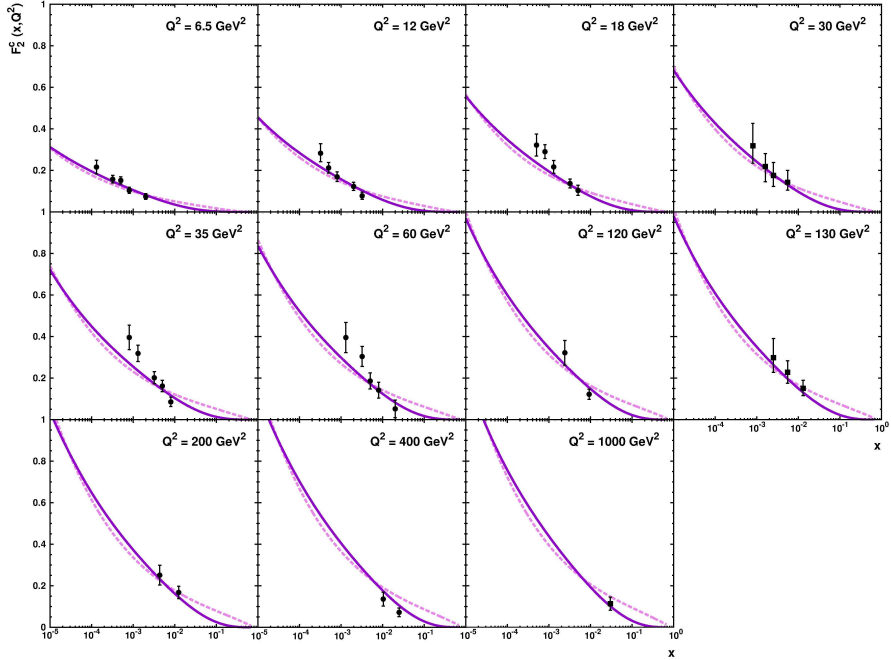}}
\caption 
{The charm structure function $F_{2c}(x,Q^2)$. 
Notation of all curves is the same as in Fig.~5. The data are taken from
\cite{H1ZEUS:2012,H1F2cb:2011,H1F2c:2010,ZEUSF2cb:2010}.} 
\label{Fig_8}
\end{figure} 
\begin{figure}[h!!]
\centerline{\includegraphics[width=0.8\textwidth]{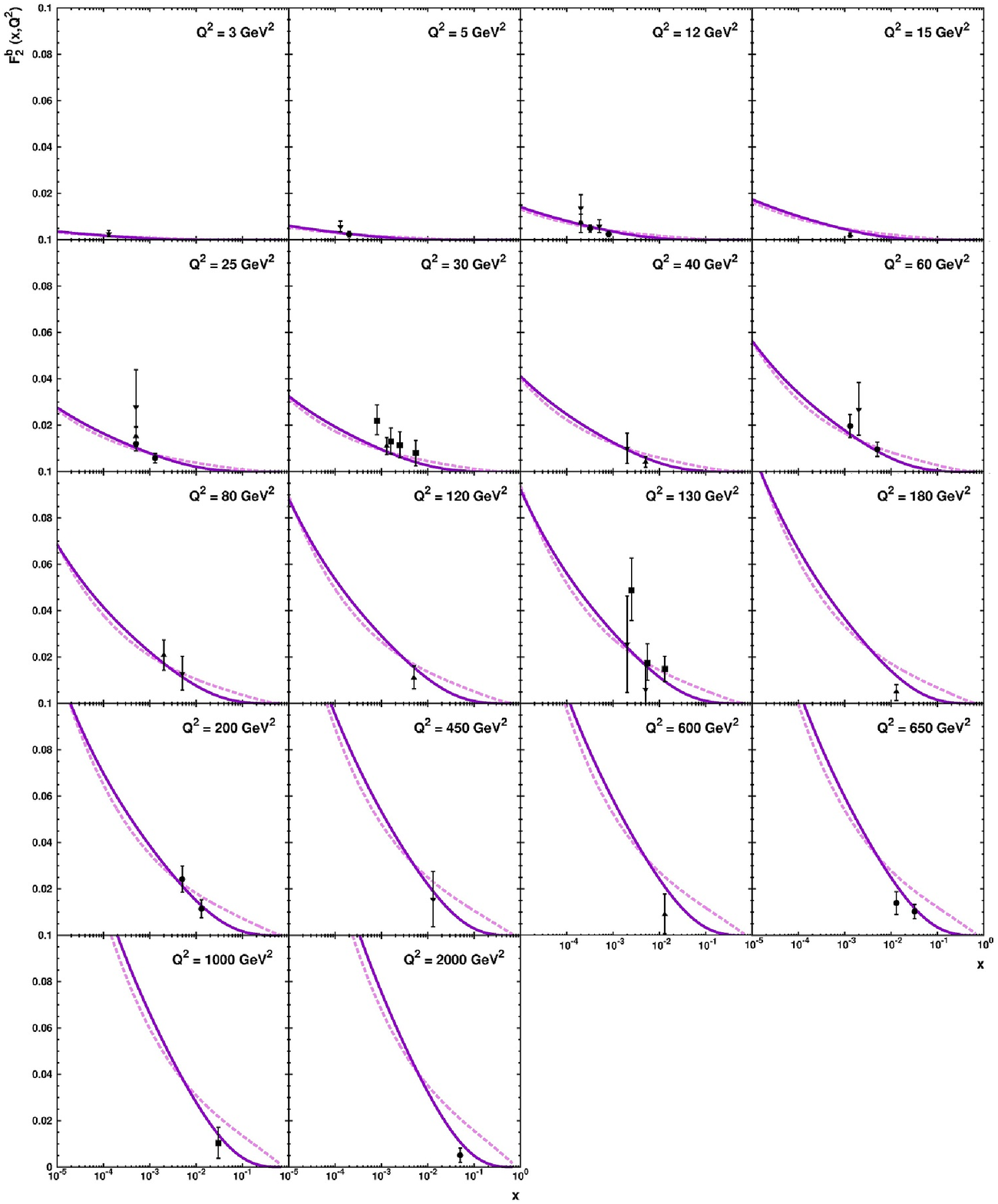}}
\caption 
{The bottom structure function $F_{2b}(x,Q^2)$. 
Notation of all curves is the same as in Fig.~5. 
The data are taken from\cite{H1ZEUS:2012,H1F2cb:2011,ZEUSF2b:2011,ZEUSF2cb:2010}.} 
\label{Fig_9}
\end{figure} 

The results of our calculations are presented in Figs.~8---10 in comparison with recent H1 and 
ZEUS data\cite{H1,H1EL,H1ZEUS:2012,H1F2cb:2011,H1F2c:2010,ZEUSF2b:2011,ZEUSF2cb:2010}.
Note that the data \cite{H1,H1EL} on 
the longitudinal SF $F_{L}(x,Q^2)$ refer to the fixed value of the hadronic mass $W = 276$~GeV.
The solid and dashed curves 
correspond to the results obtained using the modified u.g.d.~(7) and the original GBW gluon, respectively. 
One can see that the predictions obtained with the u.g.d. (7) are in a reasonable agreement with the available HERA data for 
$F_2^{c,b}(x,Q^2)$ as well
as for longitudinal SF $F_L(x,Q^2)$. Moreover, the shape of measured SF at moderate and large $x$ values at high $Q^2$ 
is better reproduced by the modified u.g.d. (7) compared to the original GBW one. When $x$ becomes small, the predictions of both gluon distributions under consideration  
practically coincide.
Therefore we conclude that the link between soft processes at the LHC and low-$x$ physics at HERA is found,  when we use
the modified u.g.d. obtained from the description of $pp$-spectra at the LHC for the analyses of the behaviour of proton structure functions at HERA. Of course, it will be important for further studies of small-$x$ physics at hadron colliders.

\section{Conclusion} 
\noindent

We have fitted  experimental data on the inclusive
spectra of charged particles produced in the central $pp$ collisions at high energies taking into account 
the sum of the quark $\rho_q$ and the gluon
contributions $\rho_g$ in (\ref{def:rhotot}) and (\ref{def:phiq}). 
The parameters of this fit do not depend on the initial energy 
in wide energy interval.
Assuming creation of soft gluons in the proton at low transverse momenta $k_t$ and 
calculating the cut one-pomeron graph between two gluons in colliding protons we found
a form for the unintegrated gluon distribution (modified u.g.d) as a function of $x$ and $k_t$ at fixed value of $Q_0^2$.
The parameters of this u.g.d. were found from the best description of the LHC data on inclusive spectra of charged hadrons produced 
in the mid-rapidity $pp$ collisions at low $p_t$. It was shown that the modified u.g.d.
is different from the original GBW u.g.d. obtained in\cite{GBW:99}
at $k_t\leq 1.6$ GeV/c and it coincides with the GBW u.g.d. at $k_t>1.6$ GeV/c.

Using the modified u.g.d. we have calculated the $q{\bar q}$ dipole-nucleon cross section ${\hat\sigma}_{\rm modif}$ 
as a function of the transverse distance $r$ between $q$ and ${\bar q}$ in the dipole and have found that it saturates  
faster than ${\hat\sigma}_{\rm GBW}$ obtained within the GBW dipole model\cite{GBW:98,GBW:99}. 
Moreover, we have shown that the relation of the modified u.g.d. and ${\hat\sigma}_{\rm modif}$ supports the form of the dipole-nucleon cross 
section and the property of 
saturation of the gluon density. 

It has been shown that the modified u.g.d. 
results in a reasonable description of the longitudinal structure function
$F_{L}(Q^2)$ at the fixed hadronic mass $W$. 
The calculations of the charm $F_2^c(x,Q^2)$ and the bottom $F_2^b(x,Q^2)$ structure functions, within the 
$k_t$-factorisation, with using the modified u.g.d. and the GBW u.g.d. show a not large difference between the results
for $F_2^c(x,Q^2)$ in the whole region of $x$ and $Q^2$ and some difference for $F_2^b(x,Q^2)$ at low $x$ and large
$Q^2$. The use of the modified u.g.d. results in a better description of the HERA data at large $Q^2$ and small $x$. 
Therefore the link between soft processes at the LHC and low-$x$ physics at HERA has been found, since 
the modification of the u.g.d. leads to a satisfactory description of both the LHC and HERA data.

\section{Acknowledgments} 
\noindent

We thank H.~Jung for extremely helpful discussions and 
recommendations in the preparation
of this paper.
The authors are grateful to S.P.~Baranov, J.~Bartels, B.I.~Ermolaev,
A.V.~Kotikov, E.A.~Kuraev, L.N.~Lipatov, E.~Levin, 
M.~Mangano, C.~Merino, M.G.~Ryskin, Yu.M.~Shabelskiy for useful discussions 
and comments. A.V.L. and N.P.Z. are very grateful to the
DESY Directorate for the support within the Moscow --- DESY project on Monte-Carlo
implementation for HERA --- LHC.
This research was also supported by the 
FASI of the Russian Federation (grant NS-3920.2012.2),
FASI state contract 02.740.11.0244 and 
RFBR grants 11-02-01454-a, 11-02-01538-a and 12-02-31030.

\end{document}